# Observation by resonant angle-resolved photoemission of a critical thickness for 2-dimensional electron gas formation in SrTiO$_3$ embedded in GdTiO$_3$

S. Nemšák[1,2,3], G. Conti[1,2], G. K. Palsson[1,2,&], C. Conlon[1,2], S. Cho[4], J. Rault[4], J. Avila[4], M.-C. Asensio[4], C. Jackson[5,♥], P. Moetakef[5,♦], A. Janotti[5], L. Bjaalie[5], B. Himmetoglu[5], C. G. Van de Walle[5], L. Balents[6], C. M. Schneider[3], S. Stemmer[5], and C. S. Fadley[1,2]

[1]Department of Physics, University of California, 1 Shields Ave, Davis, CA-95616, USA

[2]Materials Sciences Division, Lawrence Berkeley National Laboratory, 1 Cyclotron Rd, Berkeley, CA-94720, USA

[3] Peter-Grünberg-Institut PGI-6, Forschungszentrum Jülich, 52425 Jülich, Germany

[4]Synchrotron SOLEIL, L'Orme des Merisiers, Saint-Aubin, 91192 Gif sur Yvette Cedex, France

[5]Materials Department, University of California, Santa Barbara, CA-93106, USA

[6]Department of Physics, University of California, Santa Barbara, CA-93106, USA

[&]Present address: Department of Physics, Uppsala University, Uppsala, SE-751 20 Sweden

[♥]Present address: HRL Laboratories, LLC, Malibu, CA 90265 USA

[♦]Present address: Department of Chemistry, University of Maryland, College Park, MD 20742, USA

**ABSTRACT**

For certain conditions of layer thickness, the interface between GdTiO$_3$ (GTO) and SrTiO$_3$ (STO) in multilayer samples has been found to form a two-dimensional electron gas (2DEG) with very interesting properties including high mobilities and ferromagnetism. We have here studied two trilayer samples of the form [2 nm GTO/1.0 or 1.5 unit cells STO/10 nm GTO] as grown on (001) (LaAlO$_3$)$_{0.3}$(Sr$_2$AlTaO$_6$)$_{0.7}$ (LSAT), with the STO layer thicknesses being at what has been suggested is the critical thickness for 2DEG formation. We have studied these with Ti-resonant angle-resolved (ARPES) and angle-integrated photoemission and find that the spectral feature in the spectra associated with the 2DEG is present in the 1.5 unit cell sample, but not in the 1.0 unit cell sample. We also observe through core-level spectra additional states in Ti and Sr, with the strength of a low-binding-energy state for Sr being associated with the appearance of the 2DEG, and we suggest it to have an origin in final-state core-hole screening.

**INTRODUCTION**

There is presently high interest in bilayer and multilayer structures involving metal oxides, due to the novel electronic states that can develop at the interfaces between the different constituents, which can often be markedly different from the states in the native materials[1,2,3]. We focus on the detailed energy-, momentum-, and depth- resolved nature of such a 2DEG at a buried interface between the band insulator $SrTiO_3$ (STO) and the ferromagnetic Mott insulator $GdTiO_3$ (GTO).

The STO/GTO system has been studied extensively by Stemmer and co-workers[4,5,6,7,8,9,10,11,12,13], and the 2DEG here has been found to exhibit extremely high carrier densities and ferromagnetic effects, with both electrostatic and doping modulation being observed to change the carrier properties. An accepted explanation for the 2DEG creation can be found in the different oxidation state of Ti atoms in $SrTiO_3$ (4+) and $GdTiO_3$ (3+)5. The terminating layer of $TiO_2$ located at the interface of GTO and STO hence contains a mixed valence of Ti atoms, and strictly speaking from the terminating STO unit cell point of view, half of an electron per unit cell is donated from GTO. These donated d- electrons of titanium occupy in-gap states in STO and act as free electrons. Neither polar catastrophe, nor oxygen vacancies, are needed for the explanation of this electron donation, by contrast with the well-known case of STO/LAO, where vacancies are thought to be dominant[14,15]. However, beyond this, the strong role of structural distortions and symmetry breaking near the interface have been pointed out[11], theory has suggested the possibility of dimer Mott insulator (DMI) formation for the case of a single monolayer of SrO embedded in GTO[16] that could play a role for thicker STO layers as well, and conductivity has also been suggested to be due to small-polaron formation[17].

As essential background for the current study, we will refer throughout this paper to the results of a recent extensive photoemission investigation of an STO/GTO multilayer, consisting of twenty bi-layers, each comprising 6 unit cells (u.c.) of STO (nominally 23.7 Å) and 3 u.c. of GTO (nominally 11.8 Å) deposited on a (001) $(LaAlO_3)_{0.3}(Sr_2AlTaO_6)_{0.7}$ (LSAT) substrate and capped by a final STO layer (5 u.c. or 19.8 Å thick)[18]. An array of photoemission techniques was used to directly observe the 2DEG, including its energy and depth distributions and its momentum (***k***) dispersions. Core-level spectra were also studied, and found to exhibit energy shifts associated with the 2DEG. Theoretical calculations based on local-density (LDA) theory with hybrid functionals were also carried out. From this study, it has been determined that there are two sets of valence states observable in the multilayer in the STO energy gap: a lower Hubbard band (LHB) associated with GTO centered at a Fermi-referenced binding energy (BE) of about 1.0 eV and the 2DEG centered at about 0.5 eV. Both states are also found through resonant photoemission to involve a strong component of Ti 3d.

Further details of this study, including theoretical calculations that confirm the above conclusions, appear elsewhere[18].

**EXPERIMENTAL RESULTS AND DISCUSSION**

In the present investigation, we have further explored the formation of the 2DEG in STO at the STO/GTO interface for the case of trilayer samples consisting of two different thicknesses of STO near what has been found to be a critical minimum thickness for forming the 2DEG8. The two samples are illustrated in Figs. 1(a) and (b), and they consisted of a top layer of GTO of 2 nm (ca. 5 u.c.), a layer of STO either 1 u.c. or 1.5 u.c in thickness, or more precisely containing 2 or 3 layers of composition SrO, respectively, and a bottom thick layer of GTO 10 nm thick, all grown as in the multilayer study[18] on an LSAT (001) substrate using a hybrid molecular beam epitaxy approach [4,9]. Transmission electron microscopy (TEM) high-angular-aperture dark-field (HAADF) measurements confirmed the high quality of the interfaces between these layers, and images from a very similar sample are shown elsewhere[8]. The two samples are found to have vastly different transport properties, as illustrated in Fig. 2. Over the temperature range 25-300K, the sheet resistance is between 30 and 50 k$\Omega$ for the 1.5 u.c. sample, but over 300 k$\Omega$ for the 1.0 u.c. sample (Fig. 2(a)). The carrier mobility in the 1.5 u.c. sample is also very nearly constant over this range, at 0.7±0.1 cm$^2$/Vs (Fig. 2(b)), and the carrier density is also constant at ~2.5 x 10$^{14}$/cm$^2$ (Fig. 2(c)).

The photoemission measurements were carried out at the Antares beamline of the Soleil Synchrotron,[19] which is equipped with a Scienta R4000 spectrometer permitting the measurement of an angular window of ~25 degrees along the $k_y$ direction (Fig. 1(c)). No cleaning of the sample surfaces was done, so a thin layer of C+O containing contaminants was present; this we estimate from quantitative analysis of core-level spectra using the SESSA program[20] to be ca. 2-3 Å thick. As in the previously mentioned multilayer study[18], we have investigated these samples with Ti-resonant ARPES at photon energy of 465 eV (just above the Ti 2p$_{1/2}$ absorption peak), in order to enhance the emission from the states represented by the LHB and the 2DEG, which have a high degree of Ti 3d character. We have also used a higher non-resonant energy of 815 eV that permitted penetrating more deeply into the samples. For reference, the inelastic mean free paths of the valence photoelectrons in STO and GTO, which approximately represent the mean depths of emission from the samples, are estimated to be 10 Å for 465 eV, and 15 Å for 815 eV[20]. Thus, we expect to be able to see through the top GTO layer of 20 Å thickness (with approximate attenuation factors of $e^{-2} \approx 1/7$ and $e^{-1.33} \approx 1/4$), but also find that using resonant Ti excitation is crucial to exploring the STO/GTO interface properties, as also discussed in the multilayer study[18]. Core-level spectra from all elements present have also been studied to monitor differences in initial-state charge, local ground state potential and/or final-state core-hole screening state, between the two samples.

In Fig. 3, we illustrate the effects of tuning through the Ti 2p resonances on the valence-band intensities. Fig. 3(a) shows the Ti 2p x-ray absorption spectrum, which, unlike the one measured in the multilayer study mentioned above[18], is dominated by $Ti^{3+}$ contributions. For the 1.5 u.c. sample expected to show the 2DEG, a similar resonant behavior at the energies 459 eV just above the $2p_{3/2}$ peak and 465 eV just above the $2p_{1/2}$ can be found in Figs. 3(b) and 3(c), leading to a dramatic enhancement of the states in the BE region where we expect to find the LHB and the 2DEG. Both Figs. 3(b) and 3(c) have been normalized to the strong Gd 4f- dominated peak at ~10 eV BE, so as to emphasize the resonant effects on the Ti-3d dominated features at the lowest BEs. Note also in Fig. 3(b) the clear distinction of a strongly resonant feature centered at about 1.0 eV BE and extending over ca. 0.5-1.5 eV, and a weaker feature extending from the Fermi level down to about 0.3 eV. This weaker feature also resonates at a slightly higher photon energy, by about 0.5 eV, suggesting qualitatively a different Ti 3d character. The fact that the feature at 1.0 eV is much stronger than that near $E_F$, as well as its energy range, is consistent with its being the LHB, coming predominantly from the top GTO layer. This thus strongly suggests that the feature near $E_F$ is the 2DEG residing in the STO, and this assignment has been confirmed in the multilayer study[18]. In Fig. 3(d), we compare angle-integrated matrix-element density-of-states (MEWDOS) spectra for both samples, and these clearly exhibit a strong suppression of the 2DEG feature in the 1.0 u.c. spectrum, with only a hint of it remaining. This suggests an extreme sensitivity of the formation of the 2DEG to the thickness of the STO over the 1.0-1.5 u.c. range, as suggested previously[8]. Finally, in Fig. 3(e) we show *k*-resolved ARPES results on resonance for the 1.5 u.c. sample, over the BE region from $E_F$ to ~4.0 eV and for a direction that is analogous to that used in the multilayer study (Fig. 7 in ref. 18). There is a strong similarity in the energy- and momentum- dispersions seen in these two figures, with the LHB being much more smeared out in its dispersion and less observable, as seen previously[18]. The x-ray photoelectron diffraction and density-of-states profiles that are present to some degree because of phonon-induced non-direct transitions were removed from the image using standard procedures in order to enhance sharp dispersive features, such as those of 2DEG[18]. We thus conclude that the 2DEG forms in the 1.5 u.c. STO layer, but is strongly suppressed in the 1.0 u.c. STO layer, in complete agreement with the transport measurements on these two samples mentioned above[8].

We conclude with complementary core-level spectra of Ti, Gd and Sr. The Ti 2p spectra in Fig. 4(a) are similar for both samples, and we expect these to be dominated by the $Ti^{3+}$ states due to the relatively thick GTO over- and under- layers. However, for the 1.0 u.c. sample, there is a weak <u>lower-BE</u> shoulder suggestive of a different local potential, charge state, or core-hole screening state, but this is <u>not</u> in the direction expected for the $Ti^{4+}$ states in STO, which should show a higher BE. Peak fitting shows that this feature is shifted to lower BE by ~1.46 eV, and this shift is thus also different from the 2.00 eV between Ti$^{3+}$ (lower BE) and Ti$^{4+}$ (higher BE) in the Ti 2p spectrum from the multilayer[18]. The Gd 4d spectra for the two samples in Fig. 4(b) are by contrast essentially identical, indicating no change in charge, potential or screening environment for this species between the two samples. However the Sr 3d spectra in both Figs. 4(b) and 4(c) clearly indicate the presence of two components, with a separation of about 0.85 eV between them for both samples. In this case, the <u>higher-BE</u> component is more intense in the 1.0 u.c. sample, by contrast with Ti 2p. Such Sr states, separated by ca. 0.8-1.0 eV, have in fact been observed before on free STO surfaces that are not highly ordered, nor TiO$_2$ terminated[21], on partially-reduced STO surfaces treated with ion bombardment[22], and on STO surfaces etched with HF[23]. These results are significant in indicating a tendency of such a spectral feature of Sr to form in a <u>buried STO layer</u>, concomitant with an oppositely shifted feature of Ti.

We now briefly consider the possible origins of these additional Ti and Sr core-level features. Charging due to photoelectron emission can be ruled out, as it would cause all peaks to shift in the same direction and to higher binding energies. Alternatively, these peak shifts could be due to charge transfer between some of the interfacial Ti and Sr atoms, with Ti becoming more negative and Sr more positive in the 1.0 u.c. sample, as indicated by the blue arrows in Fig. 1(a), but could also be due to a more complex self-consistent potential that forms at the interface including a band offset. Beyond these considerations, strong correlation effects combined with the onset of octahedral tilts have been shown theoretically to cause a d-d band gap to open, through what has been termed a dimer Mott insulator that involves molecular dimers across two TiO layers, with an SrO layer in between[16], in analogy with the Mott-Hubbard band gap in the rare earth titanates. However, it is not clear how this would affect the Ti and Sr core levels, and the calculations have in fact been done for a single SrO layer rather than the two and three layers of our samples. Conduction in the 2DEG via a small-polaron mechanism has also been suggested in STO/GTO[17], again within a single SrO layer, but with similar results found for 2 SrO layers, and it is possible that some combination of the DMI and small polaron picture could account for the effective charge transfers seen in our core-level spectra. In the small polaron scheme, the carriers are assumed to still exist with 1 or 2 SrO layers, but to be self-localized so as to reduce mobility. This self-localization could require the kind of local potential difference that might explain our core-level spectra for Ti and Sr.

A further possible interpretation of these shifts, especially those of Sr 3d, is that the <u>low-binding energy</u> feature represents a more highly screened core hole — a final-state effect in photoemission as opposed to the ground-state effects discussed previously; and that this screening is connected with the highly delocalized 2DEG electrons, in a manner similar to the screening seen in transition metal systems when looked at with bulk sensitive hard x-ray photoemission[24]. This interpretation is supported by the relative intensities of the LBE and HBE features of the 1.0 u.c. and 1.5 u.c. samples, for which the LBE feature is weaker for 1.0 u.c. when the 2DEG is suppressed. Further support for this interpretation is given by the recent standing-wave photoemission study of an STO/GTO multilayer[18], in which the same two LBE and HBE components of Sr 3d are found to have the same standing wave profile, and thus to both be spatially distributed through the entire thickness of STO layer, rather than localized at the interfaces of it. However, the fact that the LBE Ti feature is stronger for the system in which the 2DEG is <u>suppressed</u> does not permit assigning its origin the same kind of delocalized screening effect in a simple way, unless the formation of a bandgap on the GTO side of the 2DEG suppresses the facile screening by 2DEG electrons in the STO. Further theoretical study of these shifts, e.g. using fully self-consistent LDA calculations of the full multilayer, and/or an Anderson impurity model for the screening that includes a delocalized set of electrons[24], will be necessary to better understand these intriguing core-level data.

**CONCLUSIONS**

In conclusion, we have measured core-level and valence-level photoemission spectra from two trilayer structures of GTO/STO-1.0 or 1.5 unit cell/GTO, corresponding to 2 layers of SrO or 3 layers of SrO in the STO layer, respectively, including resonant and higher non-resonant excitation relative to the Ti 2p edges. We have directly observed the 2DEG for the 1.5 u.c. sample, but not for the 1.0 u.c. sample. These results are in complete agreement with transport measurements on the same samples, and further indicate the strong sensitivity of such interface states to layer thickness.

Our core-level measurements further indicate binding-energy shifts for Ti and Sr in opposite directions that may suggest charge transfer between them, unusual potential shifts near the interface, or special final-state screening effects, which at least for Sr, appear to be directly related to the presence or absence of the 2DEG. In particular, when the low-binding-energy component of Sr is less intense, as in the 1.0 u.c. sample, the 2DEG is suppressed. An important aspect of this work is the layer configuration of the sample studied: in both cases the STO layers are buried, which naturally excludes the possibility of widely discussed surface 2DEG in STO. The sample morphology also eliminates the influence of oxygen vacancies (either in GTO or in STO) on our conclusions, as they would be expected to be present about equally in both samples and thus we would not observe the STO thickness dependence.These results further confirm the exciting prospects for controlled tuning of the electronic properties in oxide multilayer structures, as well as the utility of resonant photoemission, soft x-ray ARPES and core-level spectroscopy for studying such interface phenomena.

**FIGURE CAPTIONS:**

**Figure 1-** The sample configurations (a)-(d), together with the experimental geometry (e). In (a), the layer-by-layer configuration is shown for the 1.0 u.c. (2 SrO layer) sample, together with previously hypothesized charge transfers from $TiO_2$ layers to GdO layers (black arrows), and possible interface charge transfer to explain the Ti 2p and Sr 3d spectra (blue arrows). In (b) and (c) TEM-HAADF results at two different scales from ref. 8 are shown, verifying the high sharpness and atomic order at the interfaces in the 1.0 u.c. sample. In (d), the layer-by-layer configuration is shown for the 1.5 u.c. sample. In (e), the experimental geometry is presented, with various key angles and the photon energies defined.

**Figure 2-** Transport properties of the two trilayer samples. (a) Conductivity versus temperature for both samples. (b) Mobility for the sample with 1.5 u.c. STO (3 layers of SrO). (c) Carrier density for the sample with 1.5 u.c. STO.

**Figure 3-** Various illustrations of Ti-resonant effects in (a) x-ray absorption and (b)-(e) photoemission from the two samples. The relative photoemission intensities have been normalized to a Gd 4f-derived peak at a binding energy of 10 eV in (b)-(c), and to the lower-Hubbard-band (LHB) peak in (d). (b)-(d) are angle-integrated matrix-element-weighted density-of-states valence spectra and (e) is momentum-resolved ARPES data. (a) The Ti 2p x-ray absorption spectrum, with a pre-edge peak and the two most prominent absorption peaks due to $Ti^{3+}$ indicated. (b) Intensity of the features over the region near $E_F$ as a function of binding energy and photon energy, with the features due to the LHB and the 2DEG indicated. (c) A three-dimensional representation of the valence spectra as a function of photon energy and BE, over a broader energy range. (d) Comparison of valence spectra for the two samples on resonance. (e) A resonant (465 eV) ARPES $k_y$-Binding Energy plot from the 1.5 u.c. sample.

**Figure 4-** Core-level spectra obtained with a higher non-resonant excitation energy of 815 eV for the two samples, 1.0 u.c. STO and 1.5 u.c. STO (a) O 1s. (b) Ti 2p, with peak fitting to assess the intensity and position of the low-binding-energy shoulder on $Ti\ 2p_{3/2}$. (c) Gd 4d and Sr 3d. (d) Higher-resolution Sr 3d from the two samples, with peak fitting to resolve the high-binding-energy (HBE) and low-binding-energy (LBE) components in each.

______________________________________

# Figure 1

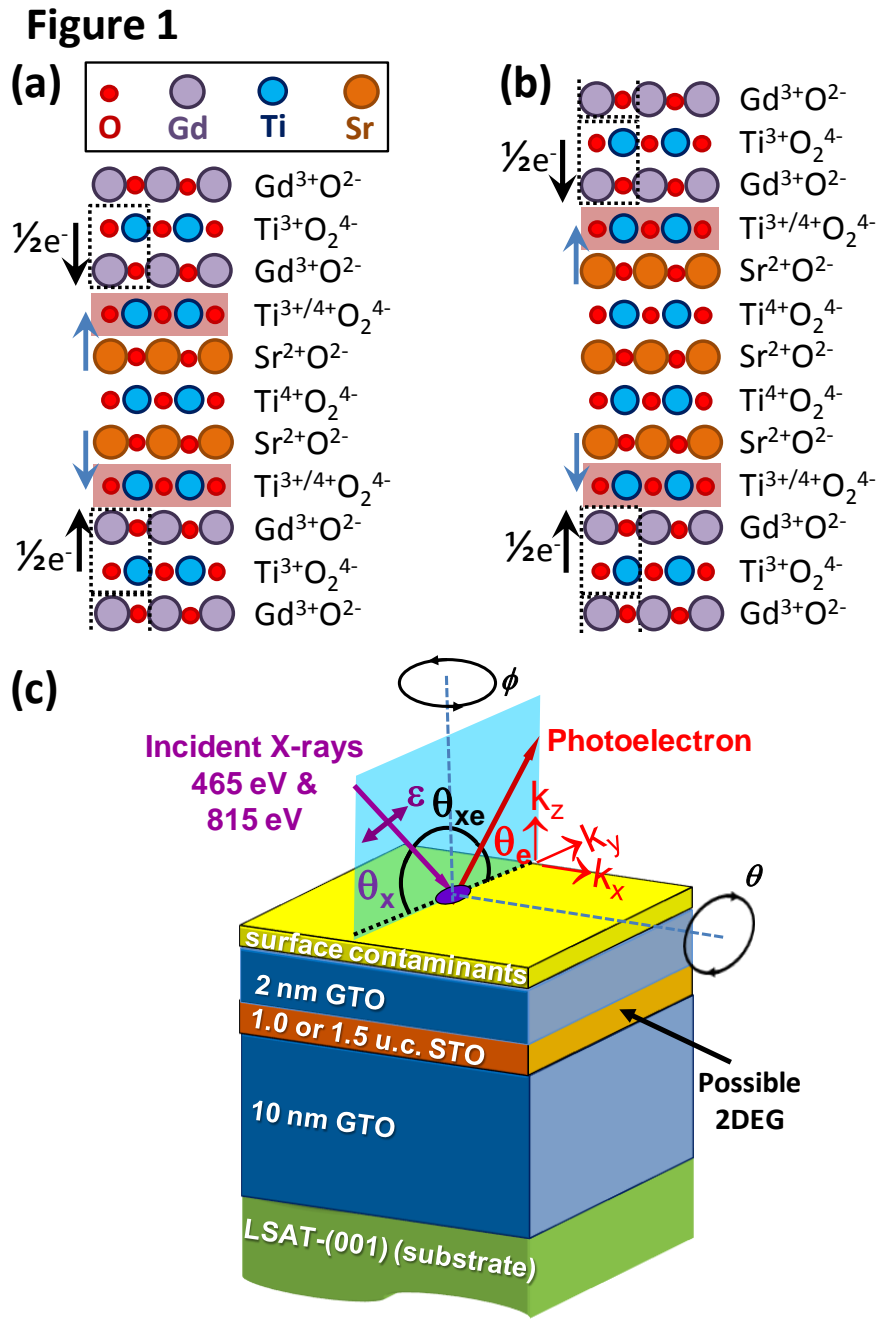

**Figure 1-** The sample configurations (a),(b), together with the experimental geometry (c). In (a), the layer-by-layer configuration is shown for the 1.0 u.c. (2 SrO layers) sample, together with previously hypothesized charge transfers from $TiO_2$ layers to GdO layers (black arrows), and possible interface charge transfer to explain the Ti 2p and Sr 3d spectra (blue arrows). In (b), the layer-by-layer configuration is shown for the 1.5 u.c. (3 SrO layers) sample. In (c), the experimental geometry is presented, with various key angles and the photon energies defined.

**Figure 2**

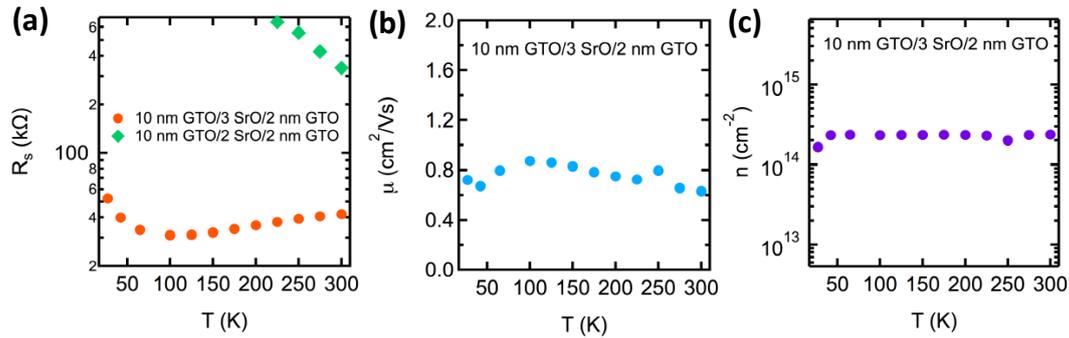

**Figure 2** - Transport properties of the two trilayer samples. **(a)** Conductivity versus temperature for both samples. **(b)** Mobility for the sample with 1.5 u.c. STO (3 layers of SrO). **(c)** Carrier density for the sample with 1.5 u.c. STO.

**Figure 3**

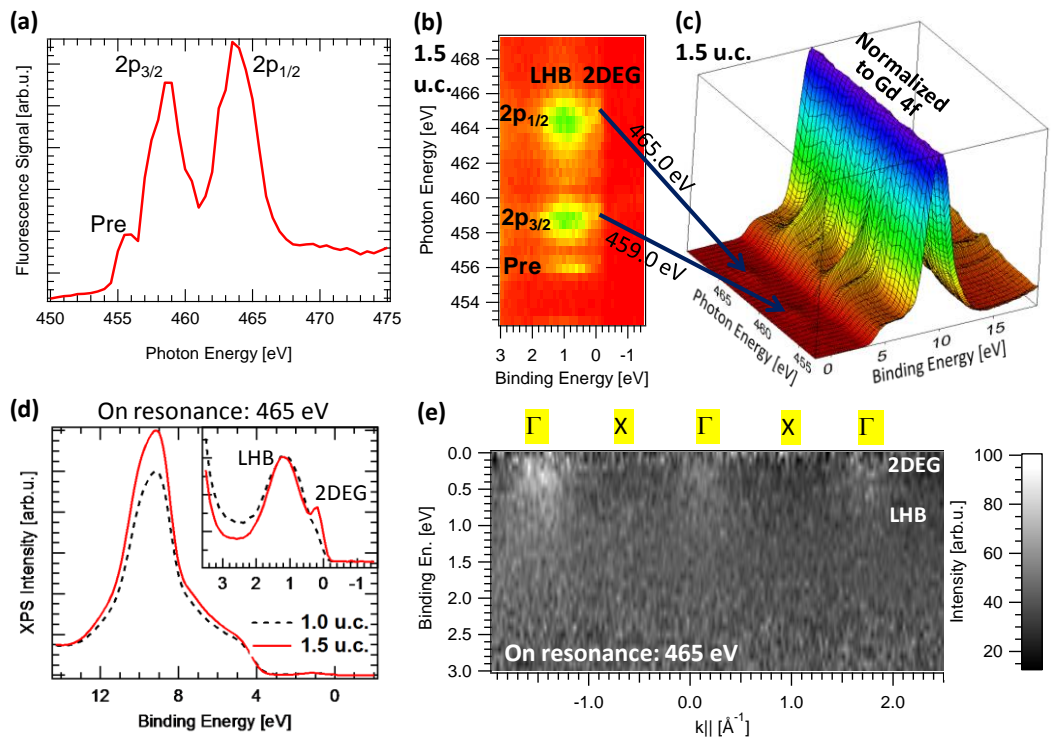

**Figure 3** - Various illustrations of Ti-resonant effects in (a) x-ray absorption and (b)-(e) photoemission from the two samples. The relative photoemission intensities have been normalized to a Gd 4f-derived peak at a binding energy of 10 eV in (b)-(c), and to the lower-Hubbard-band (LHB) peak in (d). (b)-(d) are angle-integrated matrix-element-weighted density-of-states valence spectra and (e) is momentum-resolved ARPES data. (a) The Ti 2p x-ray absorption spectrum, with a pre-edge peak and the two most prominent absorption peaks due to $Ti^{3+}$ indicated. (b) Intensity of the features over the region near $E_F$ as a function of binding energy and photon energy, with the features due to the LHB and the 2DEG indicated. (c) A three-dimensional representation of the valence spectra as a function of photon energy and BE, over a broader energy range. (d) Comparison of valence spectra for the two samples on resonance. (e) A resonant (465 eV) ARPES $k_y$-Binding Energy plot from the 1.5 u.c. sample.

**Figure 4**

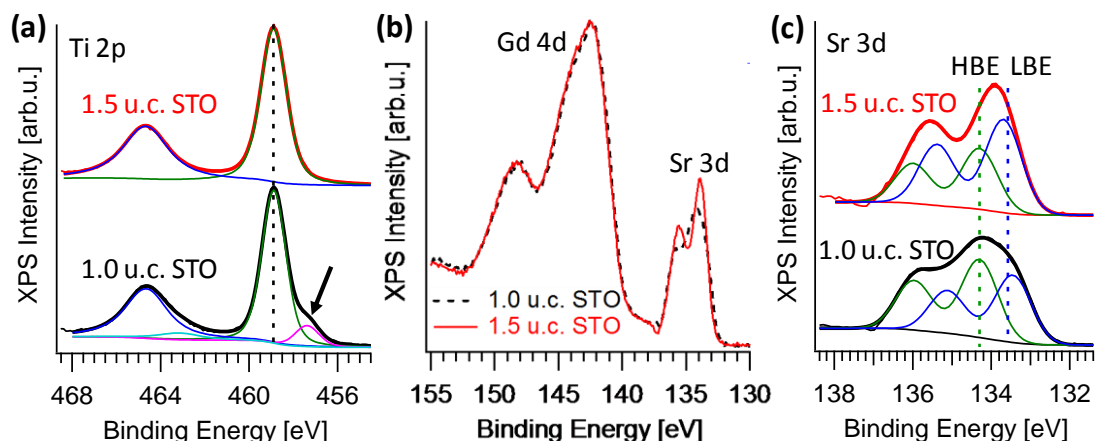

**Figure 4-** Core-level spectra obtained with a higher non-resonant excitation energy of 815 eV for the two samples, 1.0 u.c. STO and 1.5 u.c. STO. (a) Ti 2p, with peak fitting to assess the intensity and position of the shoulder on Ti $2p_{3/2}$ for 1.0 u.c. (c) Gd 4d and Sr 3d. (d) Higher-resolution Sr 3d from the two samples, with peak fitting to resolve the high-binding-energy (HBE) and low-binding-energy (LBE) components in each.

**ACKNOWLEDGEMENTS:**

Primary support for this work is from the MURI program of the Army Research Office (Grant No. W911-NF-09-1-0398). The Advanced Light Source, A.B., W.C.S., and C.S.F. are supported by the Director, Office of Science, Office of Basic Energy Sciences, Materials Sciences and Engineering Division, of the U.S. Department of Energy under Contract No. DEAC02-05CH11231. P.M. was supported by the U.S. National Science Foundation (Grant No. DMR-1006640). S.N. received support in the completion of this work from the Jülich Research Center. A.J. and C.G.V.d.W. were supported by the US Army Research Office (W911-NF-11-1-0232). L. B. was supported by the NSF MRSEC Program (DMR-1121053). Computational resources were provided XSEDE (NSF ACI-1053575). G.K.P. also thanks the Swedish Research Council for financial support. C.S.F. has also been supported during the writing of this paper by the LabEx PALM program Investissements d'Avenir"overseen by the French National Research Agency (ANR) (reference: ANR-10-LABX-0039).